\documentclass[aps,prl,twocolumn,longbibliography,superscriptaddress,final,amsmath,amssymb]{revtex4-2}
\usepackage{graphicx}
\usepackage{dcolumn}
\usepackage{bm}
\usepackage{hyperref}
\usepackage[usenames,dvipsnames]{xcolor}

\definecolor{vibrant}{HTML}{E77500}
\definecolor{muted}{HTML}{994400}

\hypersetup{
    colorlinks,
    linkcolor = vibrant,
    citecolor = muted,
    urlcolor = muted
}

\usepackage{braket}
\usepackage{enumitem}

\usepackage[english]{babel}

\begin{document}

\title{Enhanced Entanglement Gate Fidelity via Correlated and Synchronized Atom Motion}

\author{Christopher W. Wächtler}
\thanks{Both authors contributed equally to this work.}
\affiliation{Instituto de Ciencia de Materiales de Madrid ICMM-CSIC, Madrid 28049, Spain}

\author{Frederic Hummel}
\thanks{Both authors contributed equally to this work.}
\affiliation{Atom Computing, Inc., Berkeley, California, 94710, USA}

\date{\today}

\begin{abstract}
Engineered collective dissipation can correlate and even synchronize the motion of quantum systems. We show that such motional correlations can suppress motion-induced errors in two-qubit controlled-phase gates based on the Rydberg blockade in neutral-atom tweezer arrays. To leading order in the Lamb–Dicke parameter, the motional contribution to the gate infidelity is determined solely by the two-atom motional occupations and correlations, and is independent of the gate protocol. We illustrate this result for correlated thermal states and synchronized quantum van der Pol oscillators. In each case, motional correlations reduce the gate infidelity relative to independent thermal atoms at finite temperature. Our results establish atomic-motion correlations as a resource for high-fidelity neutral-atom gates and may also find applications in analog quantum simulators.
\end{abstract}

\maketitle

\textit{Introduction}---Neutral atoms in optical tweezer arrays have emerged as a leading platform for scalable quantum computing~\cite{saffmanQuantumInformationRydberg2010, kaufmanQuantumScienceOptical2021, bluvsteinQuantumProcessorBased2022, bluvsteinLogicalQuantumProcessor2024, norciaIterativeAssemblyYb2024, manetschTweezerArray61002025, chiuContinuousOperationCoherent2025}, where entangling operations are typically realized as controlled-phase (CZ) gates mediated by the Rydberg blockade. Their fidelity, however, remains a central bottleneck on the path to fault tolerance~\cite{jakschFastQuantumGates2000, levineParallelImplementationHighFidelity2019a, madjarovHighfidelityEntanglementDetection2020, everedHighfidelityParallelEntangling2023, maHighfidelityGatesMidcircuit2023, munizHighfidelityUniversalGatesYb2025, peperSpectroscopyModelingYb2025, tsaiBenchmarkingFidelityResponse2025, senooHighfidelityEntanglementCoherent2026, everedHighfidelityEntanglingGates2026}. A major error source is atomic motion~\cite{robicheauxPhotonrecoilLaserfocusingLimits2021}: the photon recoil accompanying Rydberg excitation, together with the thermal spread of the atomic wavepackets, entangles the qubit states with the motion. Moreover, heating accumulated over deep circuits aggravates this error, which in turn motivates elaborate schemes such as re-cooling during mid-circuit measurements~\cite{computingQuantumErrorCorrection2026, munizRepeatedAncillaReuse2025}.

The standard approach is to maximally cool each atom~\cite{kaufmanCoolingSingleAtom2012, thompsonCoherenceRamanSideband2013, robicheauxPhotonrecoilLaserfocusingLimits2021, janduraOptimizingRydbergGates2023, munizHighfidelityUniversalGatesYb2025, tsaiGatebasedReadoutCooling2026}. However, since for a two-qubit gate, there are two motional degrees of freedom involved, one may wonder whether  motion-correlated atoms are less prone to errors than atoms with independent stochastic motion. Realizing such correlations is by now experimentally accessible~\cite{chenScalableProgrammablePhononic2023, liuObservationSynchronizationTwo2026}. In particular, quantum synchronization~\cite{solankiQuantumSynchronization2026, schmolkeSynchronizationQuantumRegime2026, Waechtler24, cakmakSynchronizationDissipativeQuantum2026a, wachtlerSynchronizedAharonovBohmMotifs2026, zhaoQuantumSynchronizationPerturbed2025a, murtadhoCooperationCompetitionSynchronous2023g} of limit-cycle oscillators has been studied extensively in theory~\cite{leeQuantumSynchronizationQuantum2013, walterQuantumSynchronizationDriven2014g, walterQuantumSynchronizationTwo2015c, sonarSqueezingEnhancesQuantum2018b, thomasQuantumSynchronizationQuadratically2022, wachtlerTopologicalSynchronizationQuantum2023, cabotNonequilibriumTransitionDissipative2024b} and has recently been realized with a trapped ion quantum simulator~\cite{liuObservationSynchronizationTwo2026, liExperimentalRealizationSynchronization2025c}, where an engineered dissipative coupling~\cite{walterQuantumSynchronizationTwo2015c} provides full control over the synchronization of two quantum van der Pol (vdP)
oscillators~\cite{liuObservationSynchronizationTwo2026}. This raises two questions: can correlated motion improve Rydberg-gate fidelities, and does genuine synchronized atom motion offer an advantage beyond motional correlations alone?

In this Letter, we show that both correlated and synchronized atomic motion enter the gate infidelity in a form independent of the specific gate protocol to leading order of the Lamb-Dicke parameter, and reduce the infidelity for all  protocols considered here. We further study how the infidelity decrease depends on the atomic occupations and on equal versus unequal trap frequencies. Our results identify correlated atomic motion as an additional, so-far-unexplored knob for suppressing motional gate errors, complementary to, and independent of, atom cooling.
We expect contributions to entanglement gate infidelity from motion on the order of a few $10^{-4}$ and the mechanisms discussed here achieve suppression at a comparable scale, which highlights its potential as neutral atom quantum computing technologies evolve and total gate errors enter the sub-$10^{-3}$ regime.

\textit{Infidelity from atomic motion}---We consider the typical laser-driven Hamiltonian of two atoms trapped at distance $R_0$, where a laser couples the computational state $\ket{1_i}$ of atom $i=1,2$ to a Rydberg state $\ket{r_i}$. We assume that the interatomic axis is parallel to the laser axis and consider motion along this common axis. With $\hbar\equiv1$, in the rotating frame and after rotating wave approximation in the optical transition, the Hamiltonian reads
\begin{equation}
\label{eq:RydbergHamiltonian}
\begin{aligned}
     H(t) =& \sum_{i=1,2} \left[\omega_i a_i^\dagger  a_i - \Delta \ket{r_i}\bra{r_i} \right] + B( R)\ket{r_1r_2}\bra{r_1r_2}\\
& + \sum_{i=1,2} \left[\frac{\Omega(t)}{2}\ket{r_i}\bra{1_i}e^{\mathrm i\eta ( a_i^\dagger + a_i)} + \text{h.c.}\right].
\end{aligned}
\end{equation}
Here, $\Omega(t)$ describes the gate pulse, $\Delta$ the laser detuning, and $a_i$ the annihilation operator of the motional mode of atom $i$ with trap frequency $\omega_i$. We assume that the two trap frequencies differ only slightly, $\omega_1 \approx \omega_2 \approx \omega$, such that both atoms can be described by a common oscillator length $\ell = (2m\omega)^{-1/2}$ and Lamb--Dicke parameter $\eta = k\ell$, with $k$ the laser wavevector. The position of atom $i$ is then $x_i = \ell(a_i^{\vphantom\dagger} + a_i^\dagger)$, and the Rydberg interaction $B(R)$ depends on the interatomic distance $R = R_0 + x_2 - x_1$. We assume a strong blockade, $|\Omega|,|\Delta| \ll B(R_0)$, such that the doubly excited state is populated at most virtually. Furthermore, at the beginning of the gate protocol, the motional state $\ket{\Psi_0}$ factorizes from the internal state. Since Eq.~(\ref{eq:RydbergHamiltonian}) couples only $\ket{1_i}\leftrightarrow\ket{r_i}$, the propagator $ U(T) = \mathcal T\exp[-\mathrm i\int_0^T  H(t)\,dt]$ is block diagonal in the computational basis $j\in\{00,01,10,11\}$, and its diagonal blocks act on the motional degrees of freedom alone. Since our focus is the effect of correlated atom motion on the gate fidelity, we further assume that all Rydberg population returns to the computational subspace at the end of the protocol $t=T$, such that the final motional states $\ket{\Psi_j}$ are normalized, and that the phases imprinted by the ideal pulse ($\eta=0$) realize the target gate. The average gate fidelity is then determined by the overlaps of the final motional states $\ket{\Psi_j} = \bra{j} U(T)\ket{j}\ket{\Psi_0}$ of the different computational states~\cite{nielsen2002simple},
\begin{equation}
\label{eq:Foverlap}
    1- F = \frac{2}{d(d+1)} \sum_{j<j'}\left(1-\text{Re}\braket{\Psi_{j}|\Psi_{j'}}\right),
\end{equation}
with $d = 4$ throughout. The infidelity is therefore determined by the pairwise overlaps $\braket{\Psi_j|\Psi_{j'}}$.


\textit{Pure motional displacements.}---As the focus of this Letter is to isolate the effect of correlated atom motion on the gate fidelity, let us first consider an idealized setup. To this end, we assume (i) identical trap frequencies, (ii) infinite Rydberg blockade, (iii) exchange-symmetric motional states, and (iv) instantaneous population transfer, i.e., instantaneous $\pi$-pulses separated by free evolution. These assumptions are relaxed later: lifting (ii)--(iv) still yields a closed expression for the infidelity, derived in the Supplemental Material (SM)~\cite{SM}, while unequal trap frequencies are treated numerically below. 

Under assumptions (ii) and (iv), the motional wavefunction is simply displaced while keeping its original shape, i.e., $\ket{\Psi_j} = D_1(\beta_1^{(j)}) D_2(\beta_2^{(j)})\ket{\Psi_0}$ with the subscript referring to the atom $i=1,2$ and the superscript to the computational basis state. The displacement operator is defined as $D(\beta)=\exp(\beta a^\dagger-\beta^\ast a)$, where $\beta$ denotes the displacement. The recoil-induced displacements scale with the Lamb-Dicke parameter $\eta $, such that $\beta_i^{(j)}=\mathcal O(\eta)$. The overlap between the two states in the computational subspaces $j$ and $j'$ is then given by
\begin{equation}
\label{eq:OverlapCompact}
    \braket{\Psi_j|\Psi_{j'}} = \braket{\Psi_0| D_1(\Delta\beta_1) D_2(\Delta\beta_2) |\Psi_0}e^{\mathrm i \phi_{jj'}},
\end{equation}
where we have used the identity $D^\dagger(\alpha) D(\beta) = D(\beta-\alpha) \exp\left[\tfrac{1}{2}\left(\beta\alpha^\ast-\beta^\ast\alpha\right)\right]$ and defined $\Delta \beta_i = \beta_i^{(j')} - \beta_i^{(j)}$. The phase in Eq.~(\ref{eq:OverlapCompact}) is given by $\phi_{jj'} = \sum_i \text{Im}[\beta_i^{(j')} \beta_i^{(j)\ast}]$. Since in Eq.~(\ref{eq:Foverlap}) only the real part of the overlap enters, and since the leading correction of $e^{\mathrm i \phi_{jj'}}$ is purely imaginary, the phase can safely be neglected \footnote{The real part enters only at $\mathcal O(\eta^4)$.}. We are then left with $\braket{\Psi_j|\Psi_{j'}} = \braket{\Psi_0|\exp( L)|\Psi_0}$ with $L= \sum_{i}\left(\Delta\beta_i a_i^\dagger-\Delta\beta_i^\ast a_i\right) = \mathcal O(\eta)$. We now assume that the motional steady state has a $U(1)$ symmetry, $a_i \to e^{\mathrm i \varphi}a_i$, such that in the expansion of the exponential, all odd moments of $L$ vanish identically, i.e., $\braket{L} =\braket{L^3} =0$, and $\braket{\exp(L)} = 1 + \frac{1}{2}\braket{L^2} + \mathcal O(\eta^4)$. This $U(1)$ symmetry holds for all cases considered below, in particular also for synchronized quantum vdP oscillators, where the steady state is non-Gaussian due to the nonlinear two-phonon loss. 
The exchange symmetry implies equal occupations, $N \equiv  \braket{a_1^\dagger a_1} = \braket{a_2^\dagger a_2}$, and a real-valued  correlation $\chi \equiv \braket{a_1^\dagger a_2}$. 
Together with $\braket{a_i a_{i'}}= 0 = \braket{a_i^\dagger a_{i'}^\dagger}$, a further consequence of the $U(1)$ symmetry, 
this gives
\begin{equation}
\begin{aligned}
    \braket{\Psi_j|\Psi_{j'}}\approx 1 - \left(N+\tfrac{1}{2}\right)\sum_{i=1,2}|\Delta \beta_i|^2 - 2\chi \mathrm{Re}\left(\Delta\beta_1^\ast \Delta\beta_2\right).
\end{aligned}
\label{eq:overlapfinal}
\end{equation}
Inserting Eq.~(\ref{eq:overlapfinal}) into the infidelity, Eq.~(\ref{eq:Foverlap}), the motional quantities $N$ and $\chi$ factor out, while all protocol dependence is collected into two real constants:
\begin{equation}
    1-F = \frac{1}{d(d+1)}\left[\left(N+\tfrac{1}{2}\right)\mathcal{A} + 2\chi \mathcal{B}\right] \geq \frac{\mathcal A/2}{d(d+1)}
    \label{eq:1mFAB}
\end{equation}
with $\mathcal{A} \equiv \sum_{j<j'} 2\left(|\Delta\beta_1|^2 + |\Delta\beta_2|^2\right)_{jj'}$ and $  \mathcal{B} \equiv \sum_{j<j'} 2 \mathrm{Re}\left[\Delta\beta_1^\ast \Delta\beta_2\right]_{jj'}$. Equation~(\ref{eq:1mFAB}) separates the gate infidelity into a protocol-dependent part, encoded in the two real constants $\mathcal A$ and $\mathcal B$, and a motional part, encoded in $N$ and $\chi$, and holds for any two-mode motional state with $U(1)$ symmetry. Since $\mathcal A$ is a sum of non-negative terms, $\mathcal A \geq 0$ always, and the infidelity retains a floor $\mathcal A/[2d (d+1)]$ even at $N=0$ due to the zero-point fluctuations; cf. Appendix A. The correlation $\chi$, in contrast, enters through $\mathcal B$, a sum of $\text{Re}\left[\Delta\beta_1^\ast \Delta\beta_2\right]$, with no definite sign. Correlations therefore reduce the infidelity whenever $\chi \mathcal B<0$. For the positively correlated motional states considered below, this requires $\mathcal B<0$, which we find for all Rydberg-blockade gate protocols studied here.  

\textit{Relaxing assumptions (ii)--(iv)}---Equation~(\ref{eq:1mFAB}) describes instantaneous $\pi$-pulses, perfect blockade, and equal motional states for each atom. A typical gate realized through Hamiltonian~(\ref{eq:RydbergHamiltonian}) violates each of these conditions, but each violation adds a contribution that remains fixed by the pulse protocol alone (see SM~\cite{SM}). During a pulse of finite duration the atom occupies a coherent superposition of the computational and Rydberg states while it moves in the trap, such that the final motional state of a computational branch is no longer a simple displacement. Since the recoil couplings at different times do not commute in general, the overlaps additionally acquire an offset $\mathcal C$. At finite blockade, the collective Rydberg state of both atoms is light-shifted by virtual population of the doubly excited state, and the gradient of that shift exerts a force along the interatomic axis. The force modifies all protocol constants $\mathcal A, \mathcal B$, and $\mathcal C$. Furthermore, the force is odd under exchange of the two atoms, while the photon recoil is even, so the interference between the two perturbations is sensitive only to motional moments that are odd under exchange. Relaxing assumption (iii) therefore introduces the occupation imbalance $\delta n = \braket{n_2}-\braket{n_1}$, and an imaginary-valued correlation $\varkappa = \text{Im} \braket{a^\dagger_1 a_2}$.
Exchanging the atoms changes  both the sign of  the moments and the protocol constants. To leading order in $\eta$, we then find (see SM~\cite{SM})
\begin{equation}
1-F = \frac{\left(\bar N+\tfrac12\right)\mathcal A + 2\chi\mathcal B + \mathcal C + \delta n\mathcal D + 2\varkappa\mathcal E}{d(d+1)} \geq \frac{\mathcal A /2+ \mathcal C}{d(d+1)},
\label{eq:1mFgeneral}
\end{equation}
where $\bar N = (\braket{n_1} + \braket{n_2})/2$ is the mean occupation and $\chi = \text{Re}\braket{a_1^\dagger a_2}$ the real part of the correlation. The constants $\mathcal{A}$--$\mathcal E$ are properties of the protocol and are derived in the SM~\cite{SM}. The infidelity is bounded from below by the vacuum; cf. Appendix A. Unequal trap frequencies, i.e. assumption (i), affect the gate only through the motional steady state and are treated numerically below.

\begin{figure*}
    \centering
    \includegraphics[width=\linewidth]{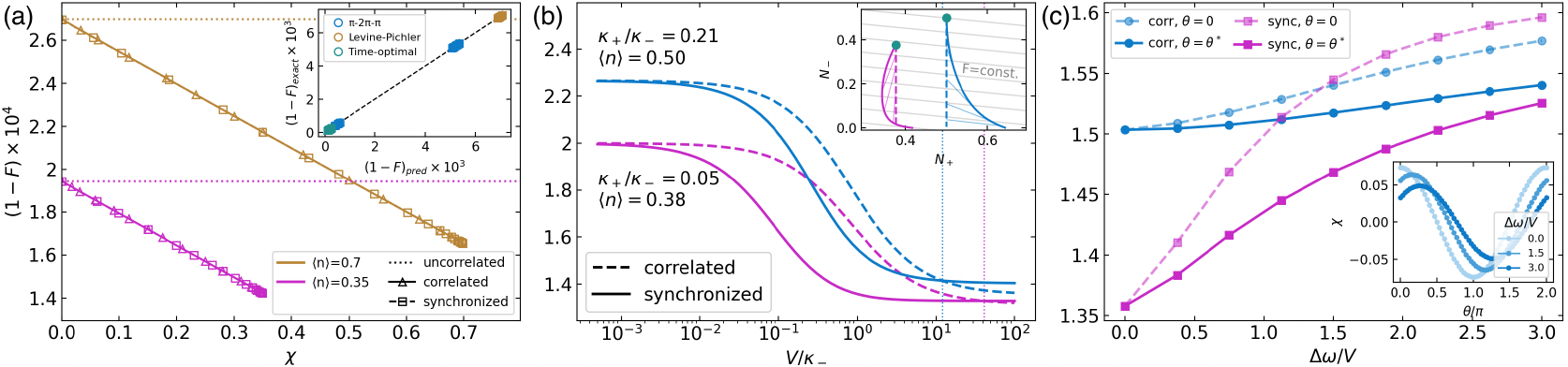}
    \caption{(a) Infidelity $1-F$ versus correlation $\chi$ at fixed occupation $\braket{n_1}=\braket{n_2}$ for the time-optimal protocol at finite Rydberg blockade  and exchange-symmetric motional states ($\delta n=\varkappa=0$). The Rydberg blockade depends on the instantaneous interatomic distance so that it fluctuates with the atomic motion. Uncorrelated states correspond to $\chi=0$.  Increasing the dissipative coupling $V$ correlates the atoms, producing the same dependence of the infidelity on $\chi$ for the correlated (triangles) and synchronized ensemble (squares). The markers correspond to full gate simulations and the lines are guides to the eye. Inset: Exact infidelity $(1-F)_\mathrm{exact}$ from full gate simulations versus the prediction $(1-F)_\mathrm{pred}$ of Eq.~(\ref{eq:1mFgeneral}) for three gate protocols at finite and infinite blockade. (b) Infidelity as a function of dissipative coupling $V$ for the time-optimal protocol and exchange-symmetric motional states. Correlated-thermal atoms (dashed) and vdP oscillators (solid) start from the same mean occupation at $V=0$. The correlated-thermal ensembles are chosen to match the occupation of the vdP oscillator at $V=0$. The vdP oscillators reach lower infidelity over most of the coupling range. Beyond the crossings (dotted vertical lines), the correlated ensemble becomes favorable. Inset: trajectories of the collective-mode occupations $N_\pm$ as $V$ is increased starting at the green dot ($V=0$). Gray lines are contours of constant $1-F$, with lower infidelity downward, while the colored lines connect points on the two trajectories at equal $V$. (c) Dependence of the infidelity on trap-frequency mismatch $\Delta\omega$. At each $\Delta\omega$ the mean occupation is held at its $\Delta\omega=0$ value $\bar N = 0.3$ by readjusting $\kappa_+$ (synchronized) or the bath occupation (correlated), with $V$ and $\kappa_-$ fixed. As $\Delta\omega$ increases, the infidelity rises toward the uncorrelated value. Dashed lines: $\theta=0$. Solid lines: $\theta=\theta^\ast=\arg\langle a_1^\dagger a_2\rangle$, which tracks the detuning-induced phase. Inset: $\chi$ versus $\theta$ for the correlated ensemble. The maximum shifts to $\theta^\ast$ as $\Delta \omega$ is increased.}
    \label{fig:ExchangeSymmetry}
\end{figure*}

\textit{Dissipatively correlated thermal atoms}---While Eqs.~(\ref{eq:1mFAB}) and (\ref{eq:1mFgeneral}) hold for any two-mode motional state with $U(1)$ symmetry, we now consider two specific models of the atomic motion. In the first, the atoms are independently damped at rate  $\gamma$ toward thermal occupations $\bar n_i$, and additionally coupled through a dissipative interaction of strength $V$. The dynamics is then described by a Lindblad equation
\begin{equation}
    \dot \varrho = \sum_{i=1,2} \mathcal L_i (\bar n_i,\gamma)\varrho + V \mathcal D[a_1 -a_2] \varrho,
    \label{eq:ThermalCorrLindblad}
\end{equation}
where $\mathcal L_i(\bar n_i, \gamma) \varrho = -\mathrm i \omega_i[a_i^\dagger a_i,\varrho]+\gamma\bar n_i \mathcal D[a_i^\dagger]\varrho + \gamma(1+\bar n_i) \mathcal D[a_i]\varrho$ are the single-atom thermalizing Liouvillians, with trap frequency $\omega_i$, and $\mathcal D[O]\varrho = O\varrho O^\dagger - \{O^\dagger O,\varrho\}/2$. For $V=0$ this describes two independent thermal atoms with steady state occupation $\braket{n_i} =\bar n_i$ and $\chi = 0$. For $V>0$, the coupling correlates the two atoms, and since all jump operators entering Eq.~(\ref{eq:ThermalCorrLindblad}) are linear in the operators, the steady state remains Gaussian and is fully characterized by the covariance matrix. Solving the resulting Lyapunov equation for equal trap frequencies $\omega_1 = \omega_2$ (see App. B) gives 
\begin{equation}
\label{eq:CommonDiffMode}
    N_+ = \frac{\bar n_1 + \bar n_2}{2},\quad N_- = \frac{ N_+\gamma}{\gamma + 2V},\quad \delta n =  \delta \bar  n \frac{\gamma}{\gamma + V}
\end{equation}
where the common and differential mode occupations are $N_\pm = \bar N \pm\chi$ and $\delta \bar n = \bar n_2 - \bar n_1$. Furthermore, $\varkappa=0$. The common mode is therefore unaffected by the dissipative coupling, while the differential mode is damped toward its ground state, which is precisely the lever for reducing the infidelity through correlated atom motion. A bath imbalance $\delta \bar n\neq 0$ adds the term $\delta n \mathcal D$, whose sign depends on the protocol and on which atom has the larger bath occupation, though this contribution is small for realistic parameters (see below).

\textit{Synchronized atomic motion: quantum vdP oscillators}---As a second model, we consider synchronized limit-cycle oscillators and ask whether genuine quantum synchronization, beyond simple motional correlations, can provide an additional fidelity advantage over the correlated-thermal case. We consider the paradigmatic model of mutual quantum synchronization: two quantum vdP oscillators coupled dissipatively~\cite{walterQuantumSynchronizationTwo2015c, liuObservationSynchronizationTwo2026}
\begin{equation}
    \dot \varrho = \sum_{i=1,2}\mathcal L_i^\mathrm{(vdP)}(\kappa_+,\kappa_-)\varrho + V \mathcal D[a_1 -a_2] \varrho,
    \label{eq:vdPLindblad}
\end{equation}
with $\mathcal L_i^\mathrm{(vdP)}(\kappa_+,\kappa_-) \varrho = -\mathrm i \omega_i[a_i^\dagger a_i,\varrho]+\kappa_+ \mathcal D[a_i^\dagger]\varrho + \kappa_- \mathcal D[a_i^2]\varrho$ the single-atom limit-cycle Liouvillians. The two-phonon loss $\mathcal D[a_i^2]$, analogous to the nonlinear dissipation of a classical Stuart--Landau oscillator, stabilizes a limit cycle with no preferred phase for each atom at $V=0$. For $V>0$, the dissipative coupling synchronizes the two phases, in direct analogy to classical synchronization~\cite{pikovskySynchronizationUniversalConcept2003, strogatzSynchronizationUniversalConcept2003}. Unlike the thermal case, there is no closed form for $\braket{n_i}$ and $\braket{a_1^\dagger a_2}$, so we determine them numerically below.

\textit{Numerical results}---We now analyze correlated atom motion for specific gate protocols. We first retain assumptions (i) and (iii), i.e., equal trap frequencies and exchange-symmetric motional states, so that $\delta n = \varkappa = 0$ in Eq.~(\ref{eq:1mFgeneral}) and  only the terms involving $\mathcal A$, $\mathcal B$, and $\mathcal C$ remain. In the following we refer to the three different motional ensembles as \emph{uncorrelated} for the case of independent thermal atomic motion, \emph{correlated} if the atoms are thermal but interact dissipatively ($V>0$) and \emph{synchronized} for the vdP oscillators. The main panel of Fig.~\ref{fig:ExchangeSymmetry}a shows the infidelity as a function of the correlation  $\chi$ at fixed mean occupation for the time-optimal protocol at finite Rydberg blockade (see App.~C). At each $V$, we adjust the single-atom dissipation ($\gamma$ and $\bar n$ for the correlated, the ratio $\kappa_+/\kappa_-$ for the synchronized ensemble) such that $\braket n$ remains fixed, which isolates the effect of the correlation. The resulting motional steady states are then used as input for full simulations of the gate protocol. Increasing the motional correlation reduces the infidelity, and both ensembles collapse onto the same line. This confirms that, to leading order in $\eta$ the motional state enters only through its second moments. Because $V$ leaves the common mode occupation $N_+$ unchanged, the latter sets a finite infidelity floor that cannot be removed by increasing the dissipative coupling $V$ alone. Thus, in Fig.~\ref{fig:ExchangeSymmetry}a, the markers accumulate and approach a finite value of $1-F$ even though $V$ is further increased.

The inset of Fig.~\ref{fig:ExchangeSymmetry}a shows the exact infidelity $(1-F)_\mathrm{exact}$ from the gate simulations versus the prediction $(1-F)_\mathrm{pred}$ of Eq.~(\ref{eq:1mFgeneral}) at $\delta n = \varkappa = 0$, for three different gate protocols (see App.~C) at finite and infinite Rydberg blockade. For each protocol, we fit $\mathcal A$, $\mathcal B$, and $\mathcal C$ to the uncorrelated and correlated ensembles only, and use the resulting constants to predict the infidelity of the synchronized ensemble without any free parameter (see SM~\cite{SM} for details). Since the synchronized steady state is non-Gaussian, its agreement with the prediction provides an independent test that, to leading order in $\eta$, the motional state enters the infidelity only through its second moments.


To leading order in $\eta$, the gate infidelity depends only on $\braket{n}$ and $\chi$. The dissipative coupling strength $V$ determines which values of $\braket{n}$ and $\chi$ the correlated and the synchronized ensemble can reach and by holding $\braket{n}$ fixed in Fig.~\ref{fig:ExchangeSymmetry}a, we removed their differences. Figure~\ref{fig:ExchangeSymmetry}b shows the infidelity as a function of $V$, with both ensembles starting from the same mean occupation $\bar n$ at $V=0$, set for the vdP oscillators by the ratio $\kappa_+/\kappa_-$. For all occupations considered, the synchronized ensemble (solid lines) reaches a lower infidelity than the correlated one (dashed lines) over a wide range of coupling strength $V$, until the curves cross and the ordering reverses. The advantage at small $V$ originates from the limit-cycle structure of the vdP oscillators. Their amplitude is stabilized by the nonlinear damping, and only the phase is free to diffuse, so that, as for classical limit-cycle oscillators, a weak coupling suffices to lock the phases of the two atoms. For thermal atoms, in contrast, amplitude and phase fluctuate alike, and the coupling must compete with the full thermal relaxation to correlate the motion. A complementary perspective is provided by the inset, which shows representative trajectories in the $(N_+,N_-)$-plane as $V$ is increased. For the correlated thermal atoms, the coupling acts only on the differential mode, so the trajectory is a vertical line at fixed $N_+$. The nonlinear dissipation of the vdP oscillators, by contrast, couples the two modes, such that both $N_-$ and $N_+$ decrease, and the trajectory reaches lower values of both at the same $V$ (colored lines connect points of equal $V$).

Relaxing assumption (iii) while retaining equal trap frequencies leaves Eq.~(\ref{eq:1mFgeneral}) valid. The exchange-asymmetric terms $\delta n \mathcal D$ and $2\varkappa\mathcal E$ are negligible for realistic interatomic spacings, whereas the exchange-symmetric force contribution (i.e.~the blockade light shift) is comparable to the errors from photon recoil and is likewise suppressed by correlations. Exchange-asymmetric motional states, such as unequal occupations of the two atoms, have correspondingly only a minor effect on the gate fidelity; see SM~\cite{SM} for more details. 
Relaxing assumption (i), in contrast, has a significant effect on the infidelity. For unequal trap frequencies, the relative phase between the atoms gets pulled away, such that at steady state, the correlation $\braket{ a_1^\dagger a_2^{\vphantom\dagger}}$ acquires a phase lag and is reduced in amplitude. To mitigate the phase lag, we introduce a relative phase $\theta$ in the dissipative coupling in Eqs.~(\ref{eq:ThermalCorrLindblad}) and (\ref{eq:vdPLindblad}) to $V\mathcal D[a_1 - e^{\mathrm i\theta}a_2]$. Figure~\ref{fig:ExchangeSymmetry}c shows the infidelity as a function of $\Delta\omega$. To isolate the effect of frequency mismatch, we fix $\bar N=0.3$ by adjusting $\kappa_+$ for the vdP oscillators and $\bar n$ for the thermal oscillators, using the same coupling $V=\kappa_-$ for both. At $\Delta\omega=0$, this corresponds to the point $V/\kappa_-=1.0$ on the $\braket{n}=0.3$ ($\kappa_+ / \kappa_- = 0.05$) curve in Fig.~\ref{fig:ExchangeSymmetry}b. For $\theta=0$, the infidelity increases  with $\Delta\omega$ for both ensembles, and the vdP oscillators eventually become less favorable than the correlated ones, despite starting from a substantially lower infidelity. However, as shown in the inset of Fig.~\ref{fig:ExchangeSymmetry}c, by tuning $\theta$ one can maximize the  correlation $\chi$, and by choosing the optimal phase $\theta^\ast$ at each $\Delta\omega$ one can suppress part of the increase in infidelity (solid lines in the main panel). With this adjustment, the synchronized ensemble retains a lower infidelity than the correlated one.


\textit{Discussion and conclusions}---In this Letter, we have shown that correlated atomic motion provides an additional route to reducing motional errors in Rydberg-blockade gates. In particular, correlating thermal atoms dissipatively already improves the fidelity, while synchronized quantum van der Pol oscillators can provide an additional advantage in regimes where the nonlinear dynamics reshapes the motional correlations. In our analysis, we deliberately isolate the roles of occupation, correlations, symmetry breaking, and trap-frequency mismatch in order to study the underlying mechanisms. In an experimental implementation, however, these quantities will generally not be independently tunable, and their interplay will depend on the specific platform and coupling scheme. A more complete optimization therefore requires a system-specific treatment. Extending our analysis beyond $U(1)$-symmetric states, e.g., to squeezed states as recently demonstrated in optical tweezers~\cite{lienhardGenerationMotionalSqueezed2025}, could reveal further ways to mitigate motional errors.

Beyond quantum gates, correlated atomic motion may also be relevant for analog quantum simulation with Rydberg tweezer arrays. Through the position dependence of the Rydberg interactions, atomic motion couples to the spin degrees of freedom, giving rise to spin-phonon coupling and distortions of the atomic configuration~\cite{magoniMolecularDynamicsRydberg2023}, which can result in bond disorder and act as a source of decoherence that limits the coherent spin dynamics~\cite{zhangMotionalDecoherenceUltracoldRydbergatom2024, dagEmergentDisorderSubballistic2025}. Such effects may be particularly relevant in frustrated systems, where small changes in the interaction geometry can influence the competition between different spin configurations. This raises the possibility that correlations or synchronization of the motional degrees of freedom could provide an additional means of controlling such effects. Whether and to what extent this can help protect the frustrated spin physics or facilitate the preparation of the corresponding target states is an interesting direction for future work.


\textit{Acknowledgments}---
The authors thank Qiming Wu, Jeffrey M. Epstein, and Johannes Agerskov for helpful comments. C.W.W. received funding from the European Union's Horizon Europe research and innovation programme under the Marie Skłodowska-Curie Actions (MSCA) grant agreement No. 101149948. Views and opinions expressed are, however, those of the author(s) only and do not necessarily reflect those of the European Union or the European Research Council. Neither the European Union nor the granting authority can be held responsible for them. 

\section*{END MATTER}
\textit{Appendix A: Lower bound on the infidelity}---Here we show that, for any motional state with $U(1)$ symmetry, the infidelity in Eq.~(\ref{eq:1mFgeneral}) is bounded from below by its value for the motional ground state. Since all Rydberg population returns to the computational subspace, the final motional states $\ket{\Psi_j}$ are normalized. With the ideal gate phases absorbed into $\ket{\Psi_j}$, each term in Eq.~(\ref{eq:Foverlap}) can therefore be written as
\begin{equation}
\label{eq:normidentity}
1-\mathrm{Re}\braket{\Psi_j|\Psi_{j'}} = \tfrac12\left(\bra{\Psi_j}-\bra{\Psi_{j'}}\right)\left(\ket{\Psi_j}-\ket{\Psi_{j'}}\right),
\end{equation}
For the ideal pulse, $\ket{\Psi_j}=\ket{\Psi_0}$ for all $j$, such that the difference $\ket{\Psi_j}-\ket{\Psi_{j'}}$ is of first order in the motional couplings. Both the photon recoil and the blockade-induced force are linear in $a_i$ and $a_i^\dagger$. At the order of Eq.~(\ref{eq:1mFgeneral}), we thus have $\ket{\Psi_j}-\ket{\Psi_{j'}} = K\ket{\Psi_0}$ with
\begin{equation}
K = \sum_{i=1,2}\left(u_i a_i^\dagger - v_i a_i\right),
\end{equation}
where the complex coefficients $u_i, v_i$ depend on $j$, $j'$, and the pulse protocol. For a $U(1)$-symmetric state, $\braket{a_i a_{i'}} = 0 = \braket{a_i^\dagger a_{i'}^\dagger}$, and normal ordering gives
\begin{equation}
\label{eq:KdagK}
\braket{\Psi_0|K^\dagger K|\Psi_0} = \sum_{i}|u_i|^2 + \braket{c_u^\dagger c_u} + \braket{c_v^\dagger c_v} \geq \sum_i |u_i|^2,
\end{equation}
with $c_u = \sum_i u_i^\ast a_i$ and $c_v = \sum_i v_i a_i$. The right-hand side is the value in the motional ground state, where $\bar N = \chi = \delta n = \varkappa = 0$. Each term in Eq.~(\ref{eq:Foverlap}), and hence the infidelity, is therefore bounded from below by its ground-state value.

\textit{Appendix B: Covariance dynamics of thermal atoms}---Since the Lindblad Eq.~(\ref{eq:ThermalCorrLindblad}) is quadratic in the bosonic operators, the second moments obey closed equations of motion, which is most easily expressed in terms of the covariance matrix $C$ with entries $C_{ij} = \braket{a_i^\dagger a_j}$. The equation of motion of the covariance matrix is given by a Lyapunov equation for $C$, 
\begin{equation}
    \dot C = A\,C + C\,A^\dagger + D,
\end{equation}
with drift and diffusion matrices (assuming equal traps here $\omega\equiv\omega_i$)
\begin{equation}
    A = -\left(\mathrm i \omega+\frac{\gamma}{2}\right)\mathbb{I} - \frac{V}{2}\begin{pmatrix} 1 & -1 \\ -1 & 1 \end{pmatrix},~
    D = \gamma \begin{pmatrix} \bar n_1 & 0 \\ 0 & \bar n_2 \end{pmatrix}
\end{equation}
Solving for $\dot C =0$ then results in the expressions
\begin{equation}
    \bar N = \frac{\bar n_1 + \bar n_2}{2}\left(\frac{\gamma+V}{\gamma + 2V}\right), \quad  \chi =\frac{\bar n_1 + \bar n_2}{2}\left(\frac{V}{\gamma + 2V}\right),
\end{equation}
and 
\begin{equation}
    \delta n = \frac{\gamma}{\gamma + V}(\bar n_2 - \bar n_1), \quad \varkappa = 0.
\end{equation}
For the common and differential modes, $a_\pm = (a_1\pm a_2)/\sqrt{2}$, given by $N_\pm = \braket{a_\pm^\dagger a_\pm} = \bar N \pm \chi$ we then find the expressions of the main text. 

\textit{Appendix C: Gate protocols}---Throughout we use the time-optimal gate protocol with finite blockade except in the inset of Fig.~\ref{fig:ExchangeSymmetry}a, where we test our predictions against two additional standard CZ protocols for  infinite and finite blockade. In all three protocols, the target is a CZ gate up to single-qubit $Z$-rotations and we summarize them below.

(1) \emph{$\pi-2\pi-\pi$ protocol}~\cite{jakschFastQuantumGates2000}: three resonant local pulses, a $\pi$-pulse on atom 1, a $2\pi$-pulse on atom 2, and a $\pi$-pulse on atom 1. The protocol has no free parameters.

(2) \emph{Levine--Pichler protocol}~\cite{levineParallelImplementationHighFidelity2019a}: two global pulses of equal duration $\tau$ and detuning $\Delta$, with a jump $\xi$ of the laser phase between them. We use the values optimized for infinite blockade without re-optimization.

(3) \emph{Time-optimal protocol}~\cite{jandura2022timeoptimal, everedHighfidelityParallelEntangling2023}: a single global pulse of duration $T$ with laser phase $\phi(t) = A\cos(\nu t - \phi_0) + \delta_0 t$. The parameters $A$, $\nu/\Omega$, $\phi_0$, $\delta_0/\Omega$, and $\Omega T$ are optimized numerically for each blockade strength.

In the absence of motion, the time-optimal protocol realizes the CZ gate to numerical precision at both finite and infinite blockade. The same holds for the $\pi$-$2\pi$-$\pi$ and Levine--Pichler protocols at infinite blockade. At finite blockade, the latter two retain a motion-independent infidelity, since their parameters are ideal only for $B\to\infty$; this contribution is absorbed into $\mathcal C$. In the main text, we focus on the time-optimal protocol.

\textit{Appendix D: Parameters}---Throughout, we use parameters for $^{171}$Yb, with values and relative scales typical of current tweezer experiments~\cite{maHighfidelityGatesMidcircuit2023, peperSpectroscopyModelingYb2025, munizHighfidelityUniversalGatesYb2025}. They are listed in Tab.~\ref{tab:Parameters}.

\begin{table}[h]
    \caption{Atomic parameters used throughout.}
    \centering
    \begin{tabular}{llc}
        \toprule
        Quantity & Symbol & Value \\\hline
        Excitation wavelength & $\lambda$ & $302$\,nm \\
        Corresponding recoil & $\omega_r/2\pi$ & $12.8$\,kHz \\
        Trap frequencies & $\omega_i/2\pi$ & $100$\,kHz \\
        Rabi frequency & $\Omega/2\pi$ & $10$\,MHz \\
        Blockade strength & $B/\Omega$ & $10$ \\
        Lamb-Dicke parameter & $\eta $ & 0.358 \\\hline
    \end{tabular}
    \label{tab:Parameters}
\end{table}

\bibliography{ref, references}

\end{document}



\title{Supplemental Material to ``Enhanced Entanglement Gate Fidelity via Correlated and Synchronized Atom Motion''}

\author{Christopher W. Wächtler}
\thanks{C.W.W. and F.H. contributed equally to this work.}
\affiliation{Instituto de Ciencia de Materiales de Madrid ICMM-CSIC, Madrid 28049, Spain}

\author{Frederic Hummel}
\thanks{C.W.W. and F.H. contributed equally to this work.}
\affiliation{Atom Computing, Inc., Berkeley, California, 94710, USA}


\maketitle

\onecolumngrid
\section{Derivation of the infidelity due to atom motion}
\label{sec:DerivationFFirstPrinciples}

In this section, we derive the motional contribution to the gate infidelity from first principles. We consider the typical laser-driven Hamiltonian of two atoms as described in Eq.~(1) of the main text. 
To derive an expression for the gate infidelity as described in Eq.~(2) of the main text, we need to evaluate the overlaps $\braket{\Psi_j|\Psi_{j'}}$. To this end, we split $\hat H(t) = \hat H_\mathrm{free}(t) + \hat V(t)$, where $\hat H_\mathrm{free}(t) = \omega\sum_i \hat a_i^\dagger \hat a_i^{\vphantom\dagger} + \hat H_0(t)$ contains the trap and the internal Hamiltonian $\hat H_0(t)$ at $\eta = 0$. Expanding the recoil factor in Eq.~(1) in powers of $\eta$, the first-order term yields
\begin{equation}
\label{eq:Vlinearized}
    \hat V(t) = \eta\sum_i \left(\hat a_i^{\vphantom\dagger}+\hat a_i^\dagger\right) \hat F_i(t),
    \qquad \hat F_i(t) = \mathrm i\frac{\Omega(t)}{2}\ket{r_i}\bra{1_i} + \text{h.c.}
\end{equation}
At second order, the expansion corrects the laser drive already present in $\hat H_0$ by a factor involving $(\hat a_i+\hat a_i^\dagger)^2$. By the $U(1)$ symmetry of $\ket{\Psi_0}$, only the diagonal contribution proportional to the motional occupation survives. It rescales the Rabi frequency (i.e.~the Debye--Waller effect) and imprints a phase that, at order $\eta^2$, does not enter $\text{Re}\braket{\Psi_j|\Psi_{j'}}$. We therefore retain only the linear recoil in $\hat V$ below. The force arising at finite blockade enters in the same way and is discussed later. 

In the interaction picture with respect to $\hat H_\mathrm{free}$, the overlap is $\braket{\Psi_j|\Psi_{j'}} = \braket{\Psi_0|\hat G_j^\dagger(T)\, \hat G_{j'}(T)|\Psi_0}$, where $\hat G_j(T)$ follow from the Dyson series to second order, i.e., 
\begin{equation}
\label{eq:Gexpansion}
\begin{aligned}
    \hat G_j(T) = 1 &- \mathrm i\eta \sum_{i}\int_0^T dt\left[\tilde a_i(t)+\tilde a_i^\dagger(t)\right] \bra{j}\tilde F_i(t)\ket{j}\\
    &- \eta^2 \sum_{i,k}\int_0^T dt'\int_0^{t'}dt\left[\tilde a_i(t')+\tilde a_i^\dagger(t')\right]\left[\tilde a_k(t)+\tilde a_k^\dagger(t)\right] \bra{j}\tilde F_i(t')\tilde F_k(t)\ket{j},
\end{aligned}
\end{equation}
where $\tilde{O}$ denotes operators in the interaction picture.
Inserting Eq.~(\ref{eq:Gexpansion}) into the overlaps, all terms of first order in $\eta$ are linear in $\hat a_i^{\vphantom\dagger}$ and $\hat a_i^\dagger$ and vanish due to the $U(1)$ symmetry. At second order,
\begin{equation}
\label{eq:MotionalCorrelator}
    \Braket{\Psi_0|\left[\tilde a_i(t')+\tilde a_i^\dagger(t')\right] \left[\tilde a_k(t)+\tilde a_k^\dagger(t)\right]|\Psi_0}
    = \begin{cases}
        (2\braket{n_i}+1)\cos(\omega\tau) - \mathrm i \sin(\omega\tau), & i=k,\\[3pt]
        2\,\text{Re}\!\left[\braket{\hat a_i^\dagger \hat a_k^{\vphantom\dagger}}e^{\mathrm i\omega\tau}\right], & i\neq k,
    \end{cases}
\end{equation}
with $\tau = t'-t$ and $\braket{n_i} = \braket{\hat a_i^\dagger \hat a_i^{\vphantom\dagger}}$. The motional state therefore enters the infidelity only through its second moments, namely the occupations $\braket{n_{1,2}}$ and the correlation $\braket{\hat a_1^\dagger \hat a_2^{\vphantom\dagger}}$, and every contribution is linear in them, with coefficients fixed by the pulse protocol alone. For the same mode, $i=k$, the $\cos(\omega\tau)$ part of the correlator yields a contribution proportional to $\braket{n_i}+\tfrac12$. The $\sin(\omega\tau)$ part stems from the commutator $[\hat a_i^{\vphantom\dagger},\hat a_i^\dagger]=1$ and yields an occupation-independent offset, which is nonzero whenever the recoil couplings at different times do not commute. For different modes, $i\neq k$, the contributions are weighted by the correlation $\braket{\hat a_1^\dagger \hat a_2^{\vphantom\dagger}}$.

\textit{Structure in each computational subspace.} For $j=00$, no atom is driven and $\hat G_{00}=1$. For $j=10$ and $j=01$, a single atom is driven, and only its own motional mode contributes. The pairs $(00,10)$ and $(00,01)$ therefore depend only on $\braket{n_1}$ and $\braket{n_2}$, respectively. In the pair $(01,10)$, the two atoms receive the same recoil kick, and the product of the two first-order terms in Eq.~(\ref{eq:Gexpansion}) couples their modes. This yields a contribution $-2\chi|\beta|^2$, where $\chi = \text{Re}\braket{\hat a_1^\dagger \hat a_2^{\vphantom\dagger}}$ and $\beta = \eta\,\omega\int_0^T dt\, e^{\mathrm i\omega t}P_r(t)$ is the recoil displacement accumulated during the pulse, with $P_r(t)$ the Rydberg population of the ideal single-atom pulse. Positive correlations therefore reduce the infidelity of this pair.

For $j=11$, both atoms are driven. The ideal drive couples $\ket{11}$ only to the bright state $\ket{W} = (\ket{1r}+\ket{r1})/\sqrt2$, while the dark state $\ket{D} = (\ket{1r}-\ket{r1})/\sqrt2$ remains unpopulated. Expressed in terms of the center-of-mass and relative modes, $\hat a_\pm = (\hat a_2\pm\hat a_1)/\sqrt2$, the recoil splits into two channels: the bright channel, $\ket{11}\leftrightarrow\ket W$, couples to $\hat a_+$, and the dark channel, $\ket{11}\leftrightarrow\ket D$, couples to $\hat a_-$. The corresponding occupations are $\braket{n_\pm} = \bar N \pm \chi$ with $\bar N = (\braket{n_1}+\braket{n_2})/2$. At finite blockade, the virtual population of $\ket{r_1r_2}$ additionally gives rise to a light shift of the bright state that depends on the interatomic distance $\hat R$. Its gradient exerts a force on the relative coordinate, which, after expanding $\hat R$ to first order in $\ell/R_0$, couples the bright-state population linearly to $\hat a_-^{\vphantom\dagger} + \hat a_-^\dagger$. Positive correlations thus increase the occupation seen by the bright channel but decrease that seen by the dark channel and the force.

The force also interferes with the bright-channel recoil. These cross terms couple the center-of-mass and relative modes and are weighted by the correlator
\begin{equation}
\label{eq:CrossCorrelator}
    \Braket{\Psi_0|\left[\tilde a_+(t')+\tilde a_+^\dagger(t')\right]\left[\tilde a_-(t)+\tilde a_-^\dagger(t)\right]|\Psi_0}
    = \delta n\,\cos(\omega\tau) - 2\varkappa\sin(\omega\tau),
\end{equation}
with the occupation imbalance $\delta n = \braket{n_2}-\braket{n_1}$ and $\varkappa = \text{Im}\braket{\hat a_1^\dagger \hat a_2^{\vphantom\dagger}}$. Under atom exchange $1\leftrightarrow2$, the moments $\bar N$ and $\chi$ are symmetric, while $\delta n$ and $\varkappa$ are antisymmetric. The recoil is exchange-symmetric, so after summing pairwise over computational states in Eq.~(2) of the main text every purely recoil-induced contribution can depend only on the symmetric moments $\bar N$ and $\chi$. The force on the other hand, acting on the relative coordinate, is exchange-antisymmetric: its second-order contributions are symmetric and involve again only $\bar N$ and $\chi$, whereas the interfering term including both recoil and force, is antisymmetric and retains $\delta n$ and $\varkappa$.

\textit{Total infidelity.} Summing the contributions of all computational subspaces in Eq.~(2), we finally obtain
\begin{equation}
\label{eq:MainResult}
    1-F = \frac{1}{d(d+1)}\left[\left(\bar N+\tfrac{1}{2}\right)\mathcal A + 2\chi\, \mathcal B + \mathcal C + \delta n\, \mathcal D + 2\varkappa\,\mathcal E\right],
\end{equation}
which separates the motional moments $\bar N$, $\chi$, $\delta n$, and $\varkappa$ from five protocol constants $\mathcal A$--$\mathcal E$. The latter depend only on the gate pulse, i.e.~its amplitudes, phase profile, and duration, and on the blockade interaction, not on the motional state. The constants $\mathcal D$ and $\mathcal E$ arise exclusively from the interference between recoil and blockade force and vanish at infinite blockade. For instantaneous pulses and infinite blockade, the motional states are rigidly displaced, $\mathcal C$ vanishes, and $\mathcal A$ and $\mathcal B$ reduce to the expressions given below Eq.~(4) of the main text. For exchange-symmetric motional states, $\delta n = \varkappa = 0$, only the terms involving $\mathcal A$, $\mathcal B$, and $\mathcal C$ remain.

\section{Details on the fitting procedure for $\mathcal A$, $\mathcal B$, and $\mathcal C$} \label{sec:Fitting}

In this section, we describe how we verify Eq.~(\ref{eq:MainResult}) numerically for the case where $\delta n = 0$ and $\varkappa=0$. To this end, we first determine the gate parameters and target phases in the absence of motion. For the $\pi$--$2\pi$--$\pi$ and Levine--Pichler protocols, we use the parameters given in App.~C of the main text. For the time-optimal protocol, we minimize the gate infidelity in the absence of motion ($\eta = 0$) separately for finite and infinite blockade. In both cases, the optimization reaches $1-F = 0$ to numerical precision. The target is a CZ gate up to single-qubit $Z$-rotations, i.e., $U = \mathrm{diag}\left(1, e^{\mathrm i\phi_1}, e^{\mathrm i\phi_2}, e^{\mathrm i(\phi_1+\phi_2+\pi)}\right)$, where the single-qubit phases $\phi_{1}$ and $\phi_2$ are read off from the propagator in the absence of motion. These target phases are kept fixed for all motional states. In actual experiments, the single-qubit phases would be calibrated on the  motional ensemble, i.e., optimized for its average phase. Here we instead use the phases of the gate without motion. Since optimizing the phases for the motional ensemble can only lower the infidelity, our values are an upper bound with respect to this choice.

We then construct the motional steady states of the uncorrelated, correlated, and synchronized ensembles at two fixed values of the mean occupation $\braket{n}$, i.e., we vary the dissipative coupling $V$ and readjust one single-atom parameter at each value of $V$. For the uncorrelated ensemble, the motional state is the product of two thermal states, $\varrho_\mathrm{th}(\braket{n})\otimes\varrho_\mathrm{th}(\braket{n})$, such that $\chi = 0$. For the correlated ensemble, the steady state of Eq.~(7) of the main text is Gaussian, and the common and differential modes are independently thermal with occupations given by Eq.~(8) of the main text. For the synchronized ensemble, we compute the steady state of Eq.~(9) of the main text numerically with $\kappa_-$ fixed, and determine $\kappa_+$ at each $V$ such that $\braket{n}$ attains its target value. In all cases, the occupation $\braket{n}$ and the correlation $\chi$ entering the analysis below are evaluated from the actual motional state.

Next, we propagate both atoms, each described by the three levels $\ket{0_i}$, $\ket{1_i}$, and $\ket{r_i}$, together with both motional modes under the Hamiltonian~(1) of the main text, where we specifically keep the dependence of the Rydberg blockade on the interatomic distance $R$. Moreover, $\Omega(t)$ and $\Delta$ contain the protocol-specific amplitudes, phases, and detunings, and the laser is tuned to the recoil-shifted resonance. The photon recoil enters through the operator $e^{\mathrm i\eta(\hat a_i^\dagger + \hat a_i)}$ without any Lamb--Dicke expansion. The motion is initialized in the full motional density matrix $\varrho_\mathrm{mot}$ of the respective ensemble, and the average gate fidelity is computed after tracing out the motion. For each protocol and blockade strength, we fit Eq.~(5) with $\delta n = \varkappa = 0$ by linear least squares, using \emph{only} the uncorrelated and correlated ensembles at both values of $\braket{n}$. With the constants fixed, we then predict the infidelity of the synchronized ensemble from Eq.~(5) without any free parameter.

As we do not use the synchronized ensemble, which is non-Gaussian, to fit any of the parameters, it provides an independent test. The agreement between the predicted and the numerically obtained infidelities directly tests whether the motional state enters the gate infidelity only through its second moments to leading order.

\section{Blockade force, exchange asymmetry, and trap-frequency mismatch}
\label{sec:Force}

In this section, we quantify the contributions to Eq.~(\ref{eq:MainResult}) that arise from the position dependence of the blockade, namely the force on the relative coordinate and, with it, the terms proportional to $\delta n$ and $\varkappa$. We then discuss how the same mechanism governs the sensitivity to unequal trap frequencies.

\textit{Blockade force.} At finite blockade, the bright state $\ket W$ is light-shifted by $\Omega^2/2B$ through the virtual population of $\ket{r_1r_2}$. The pulse parameters are calibrated to this shift at an ideal separation $R_0$. The shift, however, depends on the interatomic distance $\hat R = R_0 + \hat x_2 - \hat x_1$, and for a van der Waals interaction, $B\propto R^{-6}$, it varies by a factor $1 - 6(\hat x_2 - \hat x_1)/R_0$ to first order. Since the gate is short compared to the trap period, $\omega\ll\Omega$, the relative coordinate is frozen during the pulse, and each realization of $\hat x_2 - \hat x_1$ acts as a static miscalibration of the light shift, i.e., a detuning error on the $\ket{11}\leftrightarrow\ket W$ transition. 
In the language of Eq.~(\ref{eq:MainResult}), the force adds to $\mathcal A$ and $\mathcal B$ with $\Delta\mathcal B = -\Delta\mathcal A/2$ and leaves $\mathcal C$ unchanged to leading order in $\omega T$. It depends on the motional state only through the occupation $N_-$ of the relative mode, which is the same quantity that controls the dark recoil channel in Sec.~\ref{sec:DerivationFFirstPrinciples}. Positive correlations therefore suppress the light-shift error, and a perfectly synchronized pair, $\chi\to\bar N$, retains only the zero-point spread of the relative coordinate. For realistic parameters, the gate error due to the blockade force is comparable to the photon recoil contribution.

\textit{Exchange-asymmetric terms.} The constants $\mathcal D$ and $\mathcal E$ arise from the interference of recoil and blockade force. They are therefore bilinear in $\eta$ and $(\Omega/B)(\ell/R_0)$ and, at fixed $B/\Omega$, scale as $1/R_0$. A bath imbalance $\bar n_1\neq\bar n_2$ yields $\delta n\neq0$ and $\varkappa = 0$, see App.~B of the main text. A relative phase $\theta$ in the collective dissipator, $V\mathcal D[a_1 - e^{\mathrm i\theta}a_2]$, instead rotates the correlation: since the local rotation $a_2\to e^{\mathrm i\theta}a_2$ commutes with the single-atom Liouvillians of both ensembles, the steady state follows $\braket{\hat a_1^\dagger\hat a_2^{\vphantom\dagger}}\to e^{-\mathrm i\theta}\braket{\hat a_1^\dagger\hat a_2^{\vphantom\dagger}}$, such that $\chi\to\chi\cos\theta$ and $\varkappa\to-\chi\sin\theta$ at unchanged occupations. The infidelity then depends on $\theta$ through $2\chi(\mathcal B\cos\theta - \mathcal E\sin\theta)$, and the optimal phase, $\tan\theta^\ast = -\mathcal E/\mathcal B$, lowers the infidelity by an amount of second order in $\mathcal E/\mathcal B$. In the simulations with $B(\hat R)$ at $R_0 = 2\,\mu$m, we find $\mathcal E\approx 1.6\times10^{-5}\approx|\mathcal B|/5$, which yields a gain below $10^{-6}$ for the ensembles of Fig.~1b of the main text, while the slope of the infidelity with respect to $\delta n$ at fixed $\bar N$ is only of order $10^{-9}$. The term $\delta n\mathcal D$ is therefore negligible, and $2\varkappa\mathcal E$ contributes only if the motional state carries $\varkappa\neq0$, which requires a dissipator phase or unequal trap frequencies. A thermal imbalance alone affects the gate only through $\bar N$ and $\chi$.
The phase $\theta$ nevertheless becomes relevant for unequal trap frequencies, where it compensates a precession-induced phase lag of $\braket{\hat a_1^\dagger\hat a_2^{\vphantom\dagger}}$, see Fig.~1c of the main text and below.

\textit{Trap-frequency mismatch.} For short gate times $\Delta\omega\sim\omega\ll\Omega$, mismatched trap frequencies $\Delta\omega = \omega_1 - \omega_2$ do not affect the gate dynamics itself but only the motional steady state, so that Eq.~(\ref{eq:MainResult}) remains applicable. 
Since both the dark state coupling and the blockade force are controlled by $N_-$, these error channels grow together. Pictorially speaking, the relative atom velocity acts as a differential Doppler detuning between the atoms, which couples the bright state $\ket W$ to the dark state $\ket D$, with a variance proportional to $\eta^2\omega^2(2N_-+1)$. Compensating the phase lag by a suitable $\theta$ restores part of $\chi$ and thereby reduces $N_-$, which is the mechanism behind the partial recovery shown in Fig.~1c of the main text. With the position-dependent blockade, the phase lag also produces $\varkappa\neq0$ and thus a contribution $2\varkappa\mathcal E$ that is first order in the lag. For the parameters of Fig.~1c and $R_0=2\,\mu$m, it amounts to up to $2.5\times10^{-6}$ and is removed together with the lag by the same phase $\theta^\ast$. Optimizing $\theta$ beyond this compensation, i.e., exploiting $\mathcal E$, gains less than $10^{-6}$ of fidelity. Finally, we note that a differential polarizability of the Rydberg state would introduce a position-dependent detuning that is quadratic in $\hat x_i$. For Gaussian motional states, such a term would again enter through the second moments and be reduced for correlated motion, since its differential part involves the product of the relative and center-of-mass spreads. We omit a corresponding discussion here.